\documentclass[prl,amsmath,amssymb,superscriptaddress,twocolumn]{revtex4}
\usepackage{graphicx}

\newcommand{\bk}{{\bf k}}

\newcommand{\bB}{{\bf B}}
\newcommand{\bE}{{\bf E}}

\newcommand{\eps}{\epsilon}

\DeclareMathAlphabet{\mathpzc}{OT1}{pzc}{m}{it} 
\begin{document}
\title{Thermo-Plasma Polariton within Scaling Theory of Single-Layer Graphene}
\author{Oskar Vafek}
\address{Department of Physics, Stanford, California 94305, USA }

\date{\today}
\begin{abstract}
Electrodynamics of single-layer graphene is studied in the scaling
regime. At any finite temperature, there is a weakly damped
collective thermo-plasma polariton mode whose dispersion and
wavelength dependent damping is determined analytically. The
electric and magnetic fields associated with this mode decay
exponentially in the direction perpendicular to the graphene layer,
but unlike the surface plasma polariton modes of metals, the decay
length and the mode frequency are strongly temperature dependent.
This may lead to new ways of generation and manipulation of these
modes.
\end{abstract}
\maketitle

There has been renewed interest in the electronic properties of the
single layer graphene, spurred largely by the recent progress on the
experimental front \cite{Novoselov05,Kim05}. Most of the
experimental \cite{Novoselov05,Kim05,Kim06} and theoretical study
\cite{Gusynin05,Abanin06,CNeto06,Nomura06,Alicea06,Gusynin06} has
focused on the electronic properties in the quantum Hall regime,
where the massless relativistic dispersion of the electrons at low
energies leads to effects absent in the conventional two dimensional
electron gas.

On the other hand, the electronic properties of the single layer
graphene in the absence of the external magnetic field are arguably
as interesting as in the quantum Hall regime. The undoped system has
effectively one particle per site, but two per unit cell of a
honeycomb lattice, and in the non-interacting limit the valence and
conduction bands touch at two inequivalent K points (see e.g.
\cite{Wallace47,Wilson06}). To a very good approximation, the
dispersion near these points is $\eps_{\bk}=v_F|\bk|$ up to
$\eps\approx 1\mbox{eV}$, and the low energy excitations behave as
2D massless Dirac particles.

In the weak coupling regime, the role of the electron-electron
($e-e$) interactions at zero temperature has been studied previously
in a series of papers
\cite{Vozmediano94,Vozmediano99,Khveshchenko06}, but because the
results are scattered across the literature, it is useful to review
and interpret the main conclusions here. Since there are no mobile
carriers at $T=0$, the Coulomb interactions are not screened and the
usual Fermi liquid arguments do not guarantee sharp quasiparticle
excitations at the Fermi point. The strength of the $e-e$
interactions can be parameterized by a dimesionless coupling
constant $\alpha_F=e^2/(\eps_0\hbar v_F)$, where $\eps_0$ is the
bare dielectric constant due to the polarizability of the core
electron bands. $\alpha_F$ is of order one. Standard renormalization
group (RG) arguments show that in 2D the charge $e$ is exactly
marginal \cite{MPAF90,Herbut01}, and therefore any RG flow of
$\alpha_F$ is due to the renormalization of $v_F$. In weak coupling
the scale dependent $v_F$ grows under RG as
$v_F(k)=v^{(0)}_F+\frac{e^2}{8\pi}\log\left[\frac{k_F}{k}\right]$
and therefore $\alpha_F$ decreases, resulting in the effective
weakening of the $e-e$ interactions. The growth of $v_F$ is
eventually bounded by $c$, the speed of light, and at the eventual
infra-red fixed point the full retarded interaction between
electrons mediated by photons must be taken into account. This fixed
point is charged, with electrons moving at the speed of light, and
it is almost certainly not a Fermi liquid. Unfortunately, its
character, while fascinating, remains only of academic interest
since $c/v_F\approx 300$ and it would be impossible to probe the
system across scales differing by hundreds of orders of magnitude.
The physical properties of the system are then given by a large
crossover regime dominated by the non-retarded Coulomb interactions,
which, at least in weak coupling can be described within
perturbative RG.

The absence of the screening at $T=0$ leads to the absence of any
well defined plasma oscillations. However, as shown below, at any
{\em finite} temperature, there is a finite {\em non-activated}
density of thermally excited quasiparticles which can screen the
Coulomb interactions. Their coupling to the three dimensional
electromagnetic radiation leads to a thermo-plasma polariton mode
with frequency vanishing as $\omega_p=c|\bk|$, but crossing over
into $\omega_p=(k_BT/\hbar)\sqrt{t}(t+g)/\sqrt{t+2g}$, where
$g=e^2N\ln2/(2\pi\hbar v_F)$, $N=4$ is the number of two component
Dirac species, and $t=\hbar v_F|\bk|/(k_BT)$. The crossover
wavelength is inversely proportional to temperature and is of the
order of a millimeter at room temperature. The electromagentic
fields are confined to the two-dimensional graphene sheet within the
attenuation length $\xi_a$, which prevents the energy from radiating
outwards. The analytical expressions for $\omega_p$, $\xi_a$, and
the propagation length, $\delta\ell$, are the main results of this
paper. While surface plasma polaritons have been discussed in metals
long time ago \cite{Sarid81}, the strong temperature dependence of
the dispersion and the propagation length of the {\em thermo-plasma}
polaritons, as well as their confinement to the atomically thin
graphene sheet, may open new opportunities in the emerging field of
plasmonics \cite{Barnes03,Zia04}.

To justify the above claims, I start by solving the Maxwell
equations in the presence of a graphene sheet. The speed of light
$c$ is taken as a variable and not necessarily equal to the speed of
light in the vacuum. This allows for situations where the graphene
sheet is positioned between two dielectrics. In the scaling regime,
all energy scales can be measured in units of $k_BT$ and all lengths
can be measured in the units of thermal length $\ell_T=\hbar
v_F/(k_BT)$, where $v_F$ is the Fermi velocity at the nodal points.
As mentioned above, the strength of e-e interactions is given by the
dimensionless coupling constant, $\alpha_F$ which is the graphene
analog of the fine structure constant. Since the logarithmic
corrections to $v_F$ do not change the main physics discussed below,
for the purposes of this article these corrections will be
neglected.

Within the linear response theory the effective Lagrangian for the
electromagnetic gauge field is
\begin{eqnarray}
\mathcal{L}_{eff}=\frac{1}{2}
\Pi^{(0)}_{\mu\nu}A_{\mu}A_{\nu}+\frac{1}{2}\left[\Pi_a
a_{\mu\nu}+\Pi_b b_{\mu\nu}\right]A_{\mu}A_{\nu}|_{z=0}
\end{eqnarray}
where, $A_{\mu}|_{z=0}=\int\frac{dk_z}{2\pi}A_{\mu}(k_z)$ and for
the sake of brevity, the dependence of $A_{\mu}$ on $\omega$ and
$\bk$ was suppressed. In the Lorenz gauge the Fourier components of
the polarization tensors \cite{Vafek03} are
\begin{eqnarray}
\Pi^{(0)}_{\mu\nu}&=&
\left(-\frac{\omega^2}{c^2}+\bk^2+k_z^2\right)g_{\mu\nu}\\
a_{\mu\nu}&=&g^F_{\mu\alpha}\left(\delta_{\alpha
0}-\frac{q_{\alpha}q_0}{q^2}\right)\frac{q^2}{v_F^2\bk^2}
\left(\delta_{0\beta}-\frac{q_0 q_{\beta}}{q^2}\right)
g^F_{\beta\nu}\\
b_{\mu\nu}&=&\frac{v_F^2}{c^2}\delta_{\mu
i}\left(\delta_{ij}-\frac{k_ik_j}{\bk^2}\right)\delta_{j\nu}
\end{eqnarray}
where $\mu$ runs from $0$ to $3$ while $\alpha,\beta=0,1,2$;
$q_{\alpha}=(-i\omega,v_F\bk)$, so $q^2=-\omega^2+v_F^2\bk^2$. The
diagonal tensor $-g_{00}=g_{11}=g_{22}=g_{33}=1$ while
$ig^F_{00}=\frac{c}{v_F}g^F_{11}=\frac{c}{v_F}g^F_{22}=1$,
$g^F_{33}=0$.

In the scaling regime, the polarization functions depend only on
$\omega$, $v_F\bk$, and $T$. All the energy (length) scales can be
measured in $k_BT$ $(\ell_T)$ since the response is due to the
excitations around the nodal points, and therefore the polarization
has the scaling form
\begin{eqnarray}\label{scform}
\Pi_{a,b}=\Pi_{a,b}(\omega,v_F\bk,T)=\ell^{-1}_T\mathcal{P}_{a,b}\left(\frac{\hbar
\omega}{k_BT},|\bk|\ell_T;\alpha_F,N\right),
\end{eqnarray}
where the dependence on the dimensionless coupling constant
$\alpha_F=e^2/(\eps_0\hbar v_F)$, and the number of nodal points
$N$, was included explicitly on the right hand side. So far, the
only approximation used is the linearity of the electromagnetic
response and the existence of the scaling regime with vanishing
anomalous dimension.

The equations of motion (Maxwell's equations) for the gauge field
\begin{eqnarray}\label{eqmot}
\left[\frac{\omega^2}{c^2}-\bk^2+\frac{\partial^2}{\partial
z^2}\right]g_{\mu\nu}A_{\nu}(z) =\delta(z)\left[\Pi_a
a_{\mu\nu}+\Pi_b b_{\mu\nu}\right]A_{\nu}(z)\nonumber
\end{eqnarray}
are invariant under rotations inside the plane as well as
simultaneous scale transformation of space-time {\em and} thermal
length. Therefore in the scaling regime, any solution must satisfy
\begin{eqnarray}\label{scfreq}
\omega(\bk,k_z,T)=\frac{k_BT}{\hbar}\mathcal{W}(\ell_T|\bk|,\ell_Tk_z).
\end{eqnarray}

Clearly the z-component of $A_{\mu}$ decouples from the rest and
obeys the standard wave-equation. To proceed, we find the common
(normalized) eigenvectors of $a_{\mu\nu}$ and $b_{\mu\nu}$ and
decouple the Eq. (\ref{eqmot}). This orthogonal triad is
$e^{\parallel}_{\alpha}=\kappa_{\alpha}/|\kappa|$,
$e^{(a)}_{\alpha}=(\delta_{0\alpha}-\omega\kappa_{\alpha}/\kappa^2)|\kappa|/(c|\bk|)$,
and $e^{(b)}_{\alpha}=\eps_{0\alpha\beta3}k_{\beta}/|\bk|$, where
$\kappa_{\mu}=(\omega,c\bk)$, and $\eps_{\mu\nu\lambda\rho}$ is the
completely antisymmetric rank $4$ tensor. Projecting $A_{\mu}$ along
the four eigenvectors we have
\begin{eqnarray}
A_{\mu}=\delta_{\mu\alpha}\left[e^{\parallel}_{\alpha}A_{\parallel}
+e^{(a)}_{\alpha}A_a +e^{(b)}_{\alpha}A_b\right]+\delta_{\mu 3}A_z.
\end{eqnarray}
Note that
\begin{eqnarray}
a_{\mu\nu}A_{\mu}&=&
\left[\frac{\frac{v_F^2}{c^2}\omega^2+v_F^2\bk^2}{\omega^2-v_F^2\bk^2}\right]
e^{(a)}_{\alpha}A_a,\\
b_{\alpha\beta}A_{\beta}&=&\left[\frac{v_F^2}{c^2}\right]e^{(b)}_{\alpha}A_b.
\end{eqnarray}
The Maxwell's equations then decouple to
\begin{widetext}
\begin{eqnarray}
\ &&\left(-\frac{\omega^2}{c^2}+\bk^2-\frac{\partial^2}{\partial
z^2}\right)
\left[\frac{-\omega^2+c^2\bk^2}{\omega^2+c^2\bk^2}A_{\parallel}(z)-
\frac{2\omega c|\bk|}{\omega^2+c^2\bk^2}A_a(z) \right]=0,\\
&&\left(-\frac{\omega^2}{c^2}+\bk^2-\frac{\partial^2}{\partial
z^2}\right) \left[- \frac{2\omega
c|\bk|}{\omega^2+c^2\bk^2}A_{\parallel}(z)+
\frac{\omega^2-c^2\bk^2}{\omega^2+c^2\bk^2}A_{a}(z)
\right]+\delta(z)\left[\frac{\frac{v_F^2}{c^2}\omega^2+v_F^2\bk^2}{\omega^2-v_F^2\bk^2}\right]\Pi_a
A_a(z)=0,\\
&&\left(-\frac{\omega^2}{c^2}+\bk^2-\frac{\partial^2}{\partial
z^2}\right) A_{b}(z)+\delta(z)\left[\frac{v_F^2}{c^2}\right]
\Pi_b A_{b}(z)=0,\\
&&\left(-\frac{\omega^2}{c^2}+\bk^2-\frac{\partial^2}{\partial
z^2}\right)A_{z}(z)=0.
\end{eqnarray}
\end{widetext}
If $c^2\bk^2\neq\omega^2$ and if we let $A_{\parallel}=A_a 2\omega
c|\bk|/(c^2\bk^2-\omega^2)$, then the first equation is
automatically satisfied and the Lorenz condition
$-\frac{\omega}{c}A_0+k_{i}A_i+k_zA_z=0$ is satisfied provided that
$A_z(z)=\mbox{constant}$. Fourier transforming into the real space
along z-direction we finally have
\begin{eqnarray}
&&\left(\frac{\omega^2}{c^2}-\bk^2+\frac{\partial^2}{\partial
z^2}\right)
A_a(z)=\delta(z)\frac{\frac{v_F^2}{c^2}\omega^2-v_F^2\bk^2}{\omega^2-v_F^2\bk^2}
\Pi_a
A_a(z)\nonumber\\
&&\left(\frac{\omega^2}{c^2}-\bk^2+\frac{\partial^2}{\partial
z^2}\right) A_{b}(z)=\delta(z)\frac{v_F^2}{c^2} \Pi_b A_{b}(z).
\end{eqnarray}

The above equations resemble a Schrodinger equation for a particle
moving along a line and experiencing a $\delta-$function potential.
If the potential is attractive, we know that there is one bound
state, otherwise there are only scattering states. In this case, the
potential depends on the frequency $\omega$, on the inplane
wavevector $\bk$ and on the temperature. The bound state solutions
correspond to the evanescent electromagnetic waves whose vector
potentials $A_{a}$ and $A_{b}$ have the form
\begin{eqnarray}
A_{a,b}(\omega_{\bk}(T),\bk,z;\frac{v_F}{c})=A_{a,b}(0)\;e^{-|z|/(2\xi_{a,b})}
\end{eqnarray}
and whose inverse attenuation lengths satisfy
$\xi^{-1}_a=-\left[\left(\frac{v_F^2}{c^2}\omega^2-v_F^2\bk^2\right)/\left(\omega^2-v_F^2\bk^2\right)\right]
\Pi_a$ and $\xi^{-1}_b=-\left(v_F^2/c^2\right) \Pi_b$. The real part
of $\xi_{a,b}$ must be positive and $\omega$ and $\bk$ must satisfy
\begin{eqnarray}\label{freq}
4\xi^{2}_{a,b}\left[-\frac{\omega^2}{c^2}+\bk^2\right]=1.
\end{eqnarray}
The condition $\Re e\left[\xi_{a,b}\right]>0$ thus specifies a
region in $(\omega,\bk)$-space where the curve given by the solution
of Eq.(\ref{freq}) must lie.

Using the scaling form of the polarization functions,
Eq.(\ref{scform}), gives the scaling form of the attenuation lengths
\begin{eqnarray}
\xi_{a,b}\left(\omega,v_F\bk,T\right)=\ell_T\mathcal{X}_{a,b}\!\!\left(\frac{\hbar
\omega}{k_BT},\ell_T|\bk|;\frac{v_F}{c},\alpha_F,N\right).
\end{eqnarray}
Eq.(\ref{freq}) can then be written in terms of the dimensionless
scaling variables $s=\hbar\omega/(k_BT)$, $t=\ell_T|\bk|$ as
\begin{eqnarray}
4\mathcal{X}^2_{a,b}\!\!\left(s,t;\frac{v_F}{c},\alpha_F,N\right)\left[-\frac{v_F^2}{c^2}s^2+t^2\right]=1.
\end{eqnarray}
\begin{figure}[t]
\begin{center}
\includegraphics[width=0.5\textwidth]{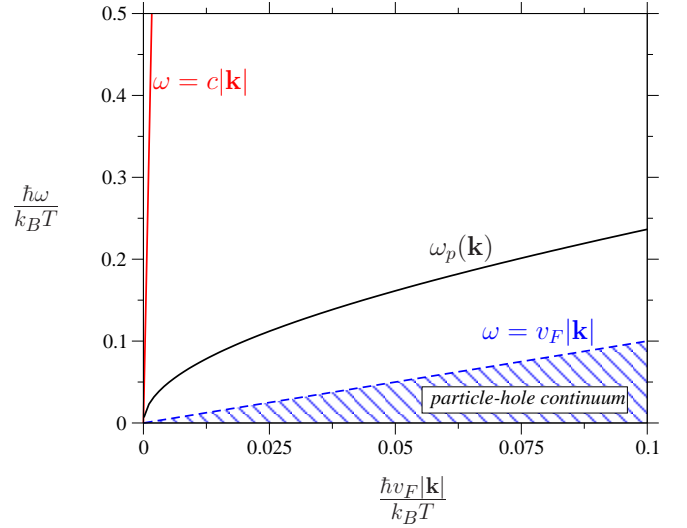}
\end{center}
\caption{(Black) The dispersion of the thermo-plasma polariton mode
in terms of the dimensionless scaling variables
$s=\hbar\omega/k_BT=\omega/(131\mbox{GHz})(\mbox{Kelvin}/T)$ and
$t=\hbar v_F|\bk|/k_BT=6.1\mu\mbox{m}|\bk|(\mbox{Kelvin}/T)$. At
small momenta, the dispersion asymptotes to $c|\bk|$- the light cone
(red). Above the crossover scale $t_c\approx
g(c-\sqrt{c^2-v_F^2})/\sqrt{c^2-v_F^2}$, the dispersion is given by
$s=\sqrt{t}(t+g)/\sqrt{t+2g}$, where the effective electron-electron
interaction strength is $g=e^2N\ln2/(2\pi\hbar v_F)$. For
orientation, at room temperature and for $|\bk|^{-1}=400\mbox{nm}$,
$\omega_p\approx6\mbox{THz}$.} \label{disp}
\end{figure}

Therefore any solution of this equation must have the scaling form,
$ s=s\left(t;\frac{v_F}{c},\alpha_F,N\right), $ which is equivalent
to Eq.(\ref{scfreq}). For purely real $\mathcal{X}_{a,b}$ this
equation sets the bound for allowed frequencies $\omega_{\bk}\leq
c|\bk|$.

To proceed, we need to resort to approximations for $\Pi_{a,b}$.
Consider first $\Pi_a$. Within the random phase approximation, we
find \cite{VafekPhd03}\cite{Khveshchenko06} that for $s^2>t^2$ the
polarization scaling function (\ref{scform})
$\mathcal{P}_a(s,t)\approx$ $
\alpha_FN\left[\frac{\ln2}{\pi}\frac{t^2-s^2}{t^2}\left(1-\frac{|s|}{\sqrt{s^2-t^2}}\right)-
i\frac{\sqrt{s^2-t^2}}{4}\tanh\frac{s}{4} \right].$ Using the above
approximation, the Eq. (\ref{freq}) can be solved by first ignoring
the imaginary part of $\Pi_a$ and then including it perturbatively.
Then, to $0^{th}$ order $
\frac{g^2}{t^4}\left[t^2-\frac{v_F^2}{c^2}s^2\right]
\left(1-\frac{|s|}{\sqrt{s^2-t^2}}\right)^2=1$, where
$g=\alpha_FN\ln2/(2\pi)$. This is a cubic equation in $s^2$ and so a
closed form solution exists. The expression is too long to present
here but the solution is plotted in Fig.\ref{disp}. In the low $\bk$
regime the dispersion follows the light cone, i.e. in the scaling
variables $s=(c/v_F)t$. At a crossover scale $t_c \approx
g(c-\sqrt{c^2-v_F^2})/\sqrt{c^2-v_F^2}$, the dispersion changes to
$s=\sqrt{t}(t+g)/\sqrt{t+2g}$. The physical origin behind these two
regimes can be understood as follows: if the interaction between the
electrons were replaced by an instantaneous Coulomb interaction, the
plasma mode would continue dispersing as
$s_{Coulomb}=\sqrt{t}(t+g)/\sqrt{t+2g}$ even for vanishing $t$.
However, at the crossover scale $t_c$, the light cone crosses
$s_{Coulomb}$ and the two mix. This is the plasmon-polariton effect.

\begin{figure}[t]
\begin{center}
\includegraphics[width=0.5\textwidth]{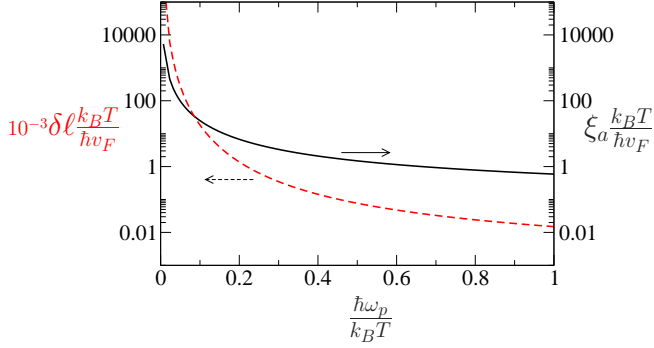}
\end{center}
\caption{The thermo-plasma polariton (in-plane) propagation length,
$\delta\ell$, (dashed red) and the (out-of-plane) attenuation
length, $\xi_a$, (solid black) normalized to thermal length
($\ell_T=\hbar v_F/(k_BT)$) vs. the mode frequency, $\omega_p$,
normalized by $k_BT/\hbar$. At room temperature and for
$\omega_p=20\mbox{THz}$, $\delta\ell\approx 2\mu\mbox{m}$.}
\label{att_prop}
\end{figure}

As $v_F/c\rightarrow 0$, the light cone regime regime shrinks as
$gv^2_F/(2c^2)$ and
\begin{eqnarray}
s\approx\sqrt{t}\frac{t+g}{\sqrt{t+2g}}-i\frac{\pi}{4\ln2}
\left[\frac{g^2t}{2g+t}\right]\tanh\left[\frac{\sqrt{t}(t+g)}{4\sqrt{t+2g}}\right].
\end{eqnarray}
The imaginary part of the frequency is the $\bk-$dependent inverse
lifetime of the monochromatic plasmon-polariton mode and has its
origin in the electron-hole scattering of the collective mode. The
distance over which the mode propagates, the propagation length
$\delta\ell$, has a scaling form as well, and it is given by the
product of the lifetime and the phase velocity $ \delta\ell=\ell_T
\Re e[s]/(t\Im m[s])$. This quantity is plotted in Fig.
(\ref{att_prop}) for $N=4$ and $v_F=8\times10^{5}m/s$. Meanwhile,
note that, within the same approximation for $\Pi_b$, there is no
weakly damped solution to Eq.(\ref{freq}).

Finally, given the scalar $A_0$ and vector potential $A_i$ solution
for the evanescent wave, the electric and magnetic field
distributions associated with the solution can be determined. While
the magnetic field is always transverse, the electric field has a
component perpendicular to the graphene sheet. Up to an overall
scale, its in-plane Fourier components are
\begin{eqnarray}
\bB&=&\left(\hat{{\bf
z}}\times\hat{\bk}\right)\frac{\omega\mbox{sgn}(z)}{2\xi_a}
\frac{\sqrt{\omega^2+c^2\bk^2}}{c^2\bk^2-\omega^2}e^{-\frac{|z|}{2\xi_a}},\\
\bE&=&\left(\frac{\hat{\bk}}{ic}(\omega^2+c^2\bk^2)+\hat{{\bf
z}}\mbox{sgn}(z)\frac{c|\bk|}{2\xi}\right)\frac{\sqrt{\omega^2+c^2\bk^2}}{c^2\bk^2-\omega^2}e^{-\frac{|z|}{2\xi_a}}.
\nonumber\\
\end{eqnarray}
The fields are thus bound to the surface which prevents the power
from radiating away from the graphene sheet.

While most of the results presented here concern ungated graphene,
the effects of applied gate voltage can be easily taken into account
within the scaling theory. Finite gate voltage $V_g$ introduces
another energy scale which must be compared to $k_BT$. If $V_g\ll
k_BT$ then the results presented above hold. If $V_g\gg k_BT$, then
the thermal energy (length) scale $k_BT$, ($\ell_T$), must be
replaced by $V_g$ or by $\hbar v_F/V_g$ respectively.

Experimentally, the temperature dependent attenuation of the EM
fields perpendicular to the layer can be measured using near field
optical microscopy, while the dispersion can be measured, for
instance, by inelastic light scattering.

There certain similarities between the {\em thermo-plasma}
polartitons discussed here and the surface plasma polaritons in thin
metals. In particular in both cases the evanescent electromagnetic
fields prevent the power from radiating outwards. However, in the
case of the surface plasma polaritons, the dispersion is essentially
fixed by the electron density and does not depend significantly on
temperature (or gate voltage). This is in sharp contrast to the
strong temperature and gate voltage dependence of the properties of
the {\em thermo-plasma} polaritons, the local manipulation of either
of which may open new ways to generate and waveguide these modes.

I wish to acknowledge useful discussions with Profs. Mark
Brongersma, Steve Kivelson and Zlatko Tesanovic. This work was
supported in part by the Stanford ITP fellowship.
\bibliography{bibliography}
\end{document}